\documentclass[%
 reprint,
superscriptaddress,
amsmath,amssymb,
aps,
pra,
]{revtex4-1}
\usepackage{graphicx}
\usepackage{dcolumn}
\usepackage{bm}
\usepackage{float}
\usepackage{multirow}
\usepackage{makecell}
\usepackage[section]{placeins}
\usepackage{float}
\usepackage{array}
\usepackage{booktabs}

\usepackage[dvipdfm,colorlinks,linkcolor=blue, urlcolor=blue, anchorcolor=blue, citecolor=blue]{hyperref}
\begin{document}


\title{Highly efficient nuclear population transfer through physics-informed neural networks}
\author{Jing Liu}%
\affiliation{College of Physics and Electronic Engineering, Northwest Normal University, Lanzhou, 730070, China}
\author{Fu-Quan Dou}
\email{doufq@nwnu.edu.cn}
\affiliation{College of Physics and Electronic Engineering, Northwest Normal University, Lanzhou, 730070, China}
\affiliation{Gansu Provincial Research Center for Basic Disciplines of Quantum Physics, Lanzhou, 730000, China}
\begin{abstract}
Nuclear coherent population transfer (NCPT) offers numerous potential applications, particularly in next-generation nuclear clocks and nuclear batteries. However, the realization of high fidelity, fast operation, and low energy consumption in NCPT remains so far challenging. Here, we employ physics-informed neural networks (PINNs) to the population transfer in an open three-level nuclear system with spontaneous emission. The method embeds the system’s control equations and boundary conditions into the loss function, thereby enabling the automatic learning of optimal laser pulse sequences that drive highly efficient population transfer. We take a short-lived excited state of $^{172}\mathrm{Yb}$ and a long-lived state of $^{229}\mathrm{Th}$ as representative examples, and systematically compare the performance of the PINNs approach with three conventional control strategies. We show that PINNs can achieve higher transfer efficiency with smaller pulse areas and shorter durations across different lifetime regimes. Our results provide a new perspective to overcome the lifetime limitation and enhance the efficiency of nuclear state transfer.
\end{abstract}
\maketitle
\section{Introduction}
Nuclear coherent population transfer (NCPT) has become an important topic in numrous fields, ranging from nuclear physics~\cite{RevModPhys.84.1177,PhysRevLett.96.142501,doi:10.1126/science.1080552,PhysRevLett.133.223001}, quantum information processing~\cite{PhysRevLett.89.207601,PhysRevA.69.052302}, to quantum computing~\cite{RevModPhys.76.1037,PhysRevC.102.064624,Yeter-Aydeniz2020,PhysRevC.105.064308}. With the continuous development of advanced x-ray free electron laser (XFEL) devices~\cite{FELDHAUS1997341,SALDIN2001357,WOOTTON2002345,ALTARELLI20112845,palffy2015straight,RevModPhys.88.015006,HUANG2021100097}, some progress has been made in theoretical studies of the interaction of the laser field with the nuclear state~\cite{PhysRevC.74.044601,di2007vacuum,PhysRevA.84.053429,Junker_2012,PhysRevLett.112.057401,PhysRevC.92.044619,10.1063/1.4935294,PhysRevLett.130.112501,PhysRevLett.133.152503}. The gap between the nuclear energy level and the laser photon energy is compensated by a moderate acceleration of the nucleus to establish resonance conditions, making the modulation of the nuclear state using coherent optical fields a reality~\cite{PhysRevC.87.054609,Liao2014,PhysRevResearch.4.L032007}. This process not only increases the controllability of nuclear quantum systems, but also provides an important opportunity to build next-generation nuclear clocks based on long-lived isomeric states~\cite{PhysRevLett.108.120802,Kazakov_2012,Seiferle2019,Peik_2021,Beeks2021,PhysRevLett.132.182501} and nuclear batteries~\cite{Walker1999,PhysRevLett.83.5242,PhysRevC.64.061302,Carroll_2004,Aprahamian2005}.

Various quantum control techniques have been successfully applied to different nuclear systems. Stimulated Raman adiabatic passage (STIRAP) is one of the most effective methods for population transfer in three-level systems~\cite{LIAO2011134}. This technique ensures that the excited state is not populated during the time evolution and is robust against fluctuations in laser parameters such as pulse intensity, duration, and time delay. However, STIRAP typically requires high-intensity laser pulses, which may be challenging to realize in practical nuclear experiments~\cite{10.1063/1.458514,RevModPhys.70.1003,VITANOV200155,vitanov2001laser}. In contrast, the $\pi$-pulse method demands smaller pulse areas but is highly sensitive to the exact resonance condition and pulse area, and leads to significant population in the excited state during the evolution~\cite{Shore01051991,PhysRevLett.96.142501,amiri2023composite}. As an alternative, the coincident-pulse technique has been proposed, in which the robustness of the system against variations in laser intensity and spontaneous emission is enhanced by increasing the number of pulse pairs~\cite{PhysRevA.85.043407,PhysRevC.94.054601}. In addition, the composite STIRAP method, proposed as an optimized version of STIRAP~\cite{PhysRevA.87.043418} has been used to population transfer in three-level nuclear systems by applying sequences of partly delayed pulse pairs with appropriate phases \cite{MANSOURZADEHASHKANI2021122119,Amiri2023}.
 Although this method reduces the peak intensity per pulse, the effect of multiple pulse pairs may result in an increased total pulse energy demand~\cite{MANSOURZADEHASHKANI2021122119,Amiri2023}.

Despite the effectiveness of the aforementioned methods under specific conditions, achieving nuclear state control that is simultaneously high-fidelity, fast, and robust remains a challenge. Recently, we apply the mixed-state inverse engineering (MIE) approach~\cite{PhysRevApplied.16.044028} to nuclear population transfer by introducing an additional laser field~\cite{PhysRevC.110.034606}. However, this process requires an additional field, which may result in increased energy consumption. Fortunately, the development of machine learning techniques, particularly physics-informed neural networks (PINNs), has opened new possibilities beyond conventional control approaches. Unlike traditional neural networks that rely on large datasets, PINNs require only the physical model itself and can provide accurate, physically consistent predictions~\cite{RAISSI2019686,karniadakis2021physics}. This enhances both interpretability and generalization capability~\cite{PhysRevLett.132.010801}. PINNs have demonstrated advantages in solving high-dimensional partial differential equations~\cite{SIRIGNANO20181339, ZENG2022111232}, many-body quantum systems~\cite{PhysRevResearch.2.033429, PhysRevResearch.5.043084}, and quantum fields~\cite{PhysRevLett.131.081601}. By minimizing a physics-based loss function, they can autonomously learn optimal control fields without requiring predefined pulse shapes. Therefore, PINNs can be applied to any quantum evolution with a well-known model~\cite{PhysRevLett.119.098301,PhysRevResearch.4.023155}, offering a new framework for addressing problems such as nuclear coherent population transfer.

Motivated by these advances, we employ the PINNs to the control of nuclear population transfer, and systematically compares its performance with three conventional quantum control techniques: coincident pulses, STIRAP and MIE. To evaluate the applicability and advantages of each method across different nuclear regimes, two representative nuclear systems are considered, characterized by short and long excited state lifetimes, respectively. Within a three-level $\Lambda$-type configuration, we construct the time-dependent evolution equations incorporating spontaneous decay and laser parameter constraints. Based on this model, we investigate the corresponding control pulses and population transfer dynamics for four control strategies, respectively. In the case of PINNs, the optimal control fields are learned using the PyTorch platform by embedding the system's dynamics directly into the loss function and optimizing toward the desired target state. Finally, a comparative analysis is performed among the four methods in terms of transfer efficiency, total pulse area, and transition time.
\begin{figure}[t]
  \centering
  \includegraphics[width=0.485\textwidth]{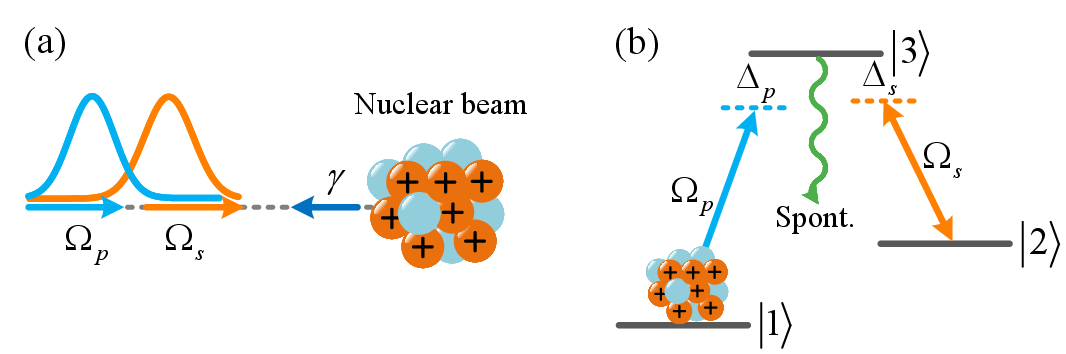}
  \caption{(a) Interaction between an accelerated nuclei and X-ray laser pulses in the laboratory frame. (b) The nuclear $\Lambda$-scheme. The initial population is in the state $|1\rangle$. The excited state $|3\rangle$ decays to other states through spontaneous emission. $\Omega_{p}$ and $\Omega_{s}$ are the pump and Stokes laser pulses, respectively. $\gamma$ is the relativistic factor. $\Delta_{p}$ and $\Delta_{s}$ denote the detunings of the corresponding lasers.}
  \label{fig1}
\end{figure}

The rest of this paper is organized as follows. In Sec.~\ref{section2}, we introduce the theoretical model and the PINN-based control scheme. In Sec.~\ref{section3}, we investigate the complete population transfer in the nuclear system and compare the performance of the PINNs approach with other methods in terms of transfer efficiency, total pulse area, and transition time. Finally, a brief discussion and summary are presented in Sec.~\ref{section4}.
\section{MODEL AND METHOD} \label{section2}
We consider the system depicted in Fig. \ref{fig1}(a), which consists of a three-state nuclear beam interacting with two x-ray laser pulses. The nuclear beam moves with velocity $v = \beta c$ and relativistic factor $\gamma = 1 / \sqrt{1 - \beta^2}$, where $\beta$ is the velocity of the nuclear particle and $c$ denoting the velocity of light in vacuum.
The interaction between the nucleus and lasers is also illustrated by a sketch in Fig. \ref{fig1}(b). States $|1\rangle$ and $|3\rangle$ are coupled by a pump laser pulse with a Rabi frequency $\Omega_p(t)$(blue arrow), while states $|2\rangle$ and $|3\rangle$ are coupled by a Stokes laser pulse with a Rabi frequency $\Omega_s(t)$(orange arrow). Spontaneous emission of the excited state $|3\rangle$ is represented by the green curve. We model the population dynamics using the density matrix approach. The density matrix $\rho(t)$ is defined as ~\cite{PhysRevC.96.044619,scully1997quantum, PhysRevC.110.034606}
\begin{eqnarray}
\rho(t)=\sum_{i,j}\rho_{ij}(t)|i\rangle\langle j|,
\end{eqnarray}
where the elements \(\rho_{ij}(t) = \langle i | \rho(t) | j \rangle\) denote the matrix components of \(\rho(t)\), with $\{i,j\}\in\{1,2,3\}$. The diagonal elements $\rho_{ii}$ represent the level populations, and off-diagonal elements $\rho_{ij}$ represent the coherences. The nuclear dynamics is governed by the master equation for the density matrix $\rho(t)$ \cite{PhysRevC.94.054601,PhysRevC.87.054609,LIAO2011134}
\begin{eqnarray}\label{nuclear master equation}
\frac{\partial}{\partial t}\rho(t)=\frac{1}{i\hbar}[H_{0}(t),\rho(t)]+\rho_{s}(t),
\end{eqnarray}
where $\hbar$ represents the reduced Planck constant. The Hamiltonian $H_{0}(t)$ reads
\begin{eqnarray}\label{Hamiltonian}
H_{0}(t)=-\frac{\hbar}{2}\begin{pmatrix}
 0 & 0 & \Omega_{p}(t)\\
 0 & -2(\Delta_{p}-\Delta_{s})& \Omega_{s}(t) \\
 \Omega_{p}(t) &\Omega_{s}(t)  &2\Delta_{p}
\end{pmatrix},
\end{eqnarray}
where $\Delta_{p(s)}$ represents the laser detuning. For simplicity, we have assumed the establishment of the full resonance condition $\Delta_{p(s)}=0$ in both theoretical and numerical calculations. The decoherence matrix $\rho_{s}(t)$ caused by spontaneous emission is given by
\begin{eqnarray}\label{decoherence matrix}
\rho_{s}(t)=\frac{\Gamma}{2}\left(\begin{array}{c c c}{{2B_{31}\rho_{33}}}&{{0}}&{{-\rho_{13}}}\\ {{0}}&{{2B_{32}\rho_{33}}}&{{-\rho_{23}}}\\ {{-\rho_{31}}}&{{-\rho_{32}}}&{{-2\rho_{33}}}\end{array}\right),
\end{eqnarray}
where $\Gamma$ is the linewidth of the excited state $|3\rangle$, and $B_{3i}$ denotes the branching ratio of the transition $|3\rangle \rightarrow |i\rangle$ $(i=1,2)$.

\begin{figure}[t]
  \centering
  \includegraphics[width=0.485\textwidth]{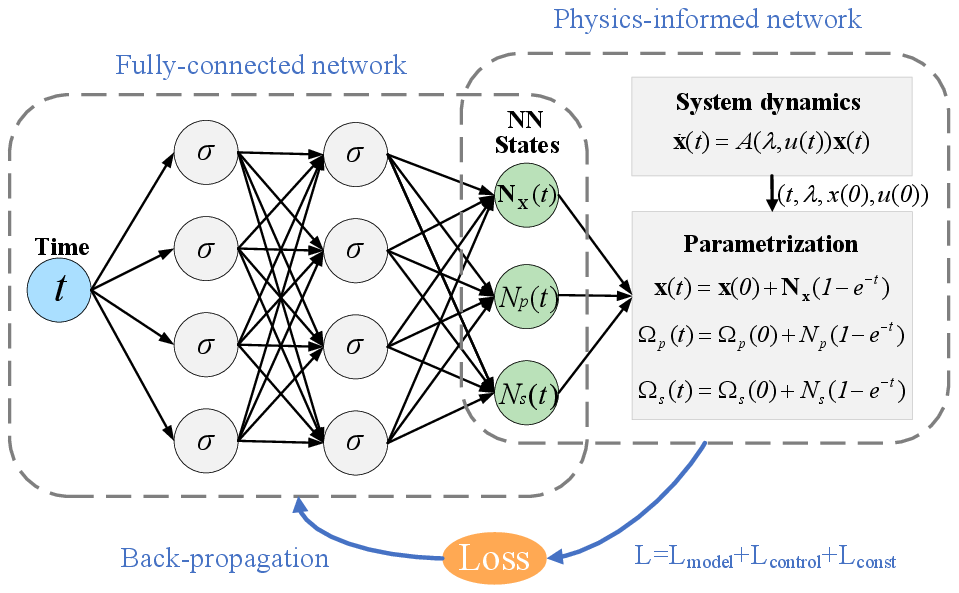}
  \caption{PINNs architecture for solving NCPT. The quantum dynamics is formulated as the differential equation $\dot{\mathbf{x}}(t) = A(\lambda, u(t)) \mathbf{x}(t)$, where $\mathbf{x}(t)$ and $u(t)$ are the state and control vectors, respectively, and $\lambda$ is the system parameters. The neural network takes discrete time vector $t$ (blue circle) as input and outputs the $\mathbf{N}_\mathbf{x}(t)$, $N_p(t)$ and $N_s(t)$ (green circle). By minimizing the loss function $\mathit{L}$, the neural network outputs the parameterized solutions for $\mathbf{x}(t)$, $\Omega_p(t)$ and $\Omega_s(t)$.}
  \label{fig2}
\end{figure}

To achieve efficient NCPT, we adopt the PINNs framework, which integrates physical laws directly into the training of neural networks. By minimizing loss functions, the framework generates smooth control functions that ensure the system evolves accurately from the initial state $\mathbf{x}_0$ to the desired target state $\mathbf{x}_d$~\cite{10.1038/s42254-021-00314-5}. As shown in Fig.~\ref{fig2}, we construct a fully connected neural network that takes time $t$ as input and outputs neural representations of both the state and the control variables. These outputs are then embedded into parameterized forms that satisfy physical constraints such as initial conditions and smooth time dependence. In this framework, the network predictions are substituted into the system dynamical equations, and deviations from the physical model are penalized through loss functions. This allows the network to learn from the underlying physics, providing an efficient and interpretable approach for optimizing control pulses $\Omega_p(t)$ and $\Omega_s(t)$.

We rewrite the nuclear state dynamics into a nonautonomous linear differential equation
\begin{equation}
\dot{\mathbf{x}}(t) = A(\lambda, u(t)) \mathbf{x}(t),
\quad \mathbf{x}(0) = \mathbf{x}_0.
\label{eq:nonautonomous_system}
\end{equation}
Here, the real state vector $\mathbf{x} =$ $(\rho_{11},$  $\rho_{22},$  $\rho_{33},$  $\mathrm{Re}[\rho_{12}],$ $\mathrm{Im}[\rho_{12}],$ $\mathrm{Re}[\rho_{13}],$ $\mathrm{Im}[\rho_{13}]$, $\mathrm{Re}[\rho_{23}],$ $\mathrm{Im}[\rho_{23}])^T$. The \(u(t) = \{\Omega_{p}(t), \Omega_{s}(t)\}\) denotes the time-dependent control variables to be optimized by the PINNs, with $\Omega_{p}(0)=0$ and $\Omega_{s}(0)=0$. The system parameters \(\lambda = \{B_{31}, B_{32}, \Gamma\}\)  include branching ratios and  linewidth of the excited. The system matrix \(A(\lambda, u(t))\) is a real \(9 \times 9\) matrix that depends on both the control variables and the parameter set. By recasting the quantum master equation Eq.~(\ref{nuclear master equation}) into the form of Eq.~(\ref{eq:nonautonomous_system}), the dynamics can be directly incorporated into the PINNs framework. Further details on the construction of the matrix $A(\lambda, u(t)) \mathbf{x}(t)$ are provided in Appendix \ref{appendix1}.

By the universal approximation theorem~\cite{HORNIK1991251}, a feed forward neural network can approximate any vector function \(F \colon \mathbb{R}^p \to \mathbb{R}^q \), where $p = 1$ and $q = 11$, corresponding to the $9$-dimensional state vector and $2$ control variables. We adopt the architecture illustrated in Fig.~\ref{fig2}, where the multilayer perceptron consists of five fully connected hidden layers, each containing $128$ neurons and activation functions denoted by $\sigma$, where $\sigma $ is $\tanh(\cdot)$, to maintain both nonlinearity and gradient stability~\cite{DUBEY202292}. We use a $9$-dimensional equally distributed time array \( t = (t_1, \dots, t_9) \) as the input to the neural network. Since PINNs do not require a structured mesh, in fact the time points \(t_i\) can be arbitrarily discretized within the time range from $t_0=0$ to $t_f$~\cite{PhysRevLett.132.010801}. The outputs of the neural network are the predicted state variable $\mathbf{N}_\mathbf{x}(t) \in \mathbb{R}^9$ and the control variables $N_p$ and $N_s$. These variables satisfy the following constraints:
\begin{eqnarray}\label{MLP}
\mathbf{x}(t) &=& \mathbf{x}(0)  + \mathbf{N}_\mathbf{x} \left(1 - e^{-t}\right), \nonumber \\
\Omega_p(t)   &=& \Omega_p(0)  +N_p \left(1 - e^{-t}\right), \nonumber \\
\Omega_s(t)   &=& \Omega_s(0)  +N_s \left(1 - e^{-t}\right),
\end{eqnarray}
where $\mathbf{x}(t)$ is the predicted state distribution equal to the initial value plus a correction term for the PINNs network prediction, avoiding numerical errors in the initial conditions by using the exponential envelop $1 - e^{-t}$ form~\cite{PhysRevE.105.065305}.

The physical information loss function defined as follows
\begin{eqnarray}\label{eq:L_total}
L = L_{\text{model}} + L_{\text{control}} + L_{\text{const}},
\end{eqnarray}
in which the component $L_{\text{model}}$ describes the system dynamics
\begin{equation}\label{eq:L_model}
L_{\text{model}} = \eta_1 \sum_{i=1}^{9} \left\| \dot{\mathbf{x}}(t_i) - A(\lambda, u(t_i)) \, \mathbf{x}(t_i) \right\|^2,
\end{equation}
where $\dot{\mathbf{x}}(t_i)$ represents the time derivative of the state vector predicted by the neural network (NN), computed using automatic differentiation provided by PyTorch package~\cite{paszke2019py}. The derivative is then compared with the time derivative obtained through the system dynamics by an $l_2$ error calculation. The magnitude of the $l_2$-norm reflects how well the network's output conforms to the underlying physical laws. By minimizing this function, the state vector will approximately satisfy the system dynamics, thus adhering to the underlying physics.
The second term on the righthand side of Eq.~(\ref{eq:L_total}) represents the control, which reads
\begin{equation}
L_{\text{control}} = \eta_2 \sum_{i=1}^{9} \left\| \mathbf{x}(t_i) - \mathbf{x}_d \right\|^2.
\label{eq:L_control}
\end{equation}
This loss term penalizes the deviation between the system's current state $\mathbf{x}(t_i)$ and the target state $\mathbf{x}_d$, thereby guiding the neural network to learn control fields that drive the system toward the target state.
The third term on the righthand side of Eq. (\ref{eq:L_total}) accounts for physical constraints
\begin{equation}
L_{\text{const}} = \eta_3 \| \rho_{11} + \rho_{33}  \|^2,
\end{equation}
where the factor \(\eta_i \, (i=1,2,3)\) regulates the relevance of each component compared to the total loss \(L\), and satisfies \(0 \leq \eta_i \leq 1\). The $\rho_{11}$ and $\rho_{33}$ respectively represent the population of the ground and intermediate states, which are minimized to approach zero as closely as possible in order to avoid physically meaningless solutions.

The PINNs is trained using the Adam optimizer with learning rate \(2 \times 10^{-4}\) and training spans $30,000$ epochs. The weights for \(L_{\text{model}}\), \(L_{\text{control}}\), and \(L_{\text{const}}\) are \(\eta_1 = 0.0899\), \(\eta_2 = 0.040\), and \(\eta_3 = 0.800\), respectively, and are manually tuned. The network parameters are initialized randomly and updated via backpropagation. At each epoch, the loss components are computed, and the total loss is minimized. Throughout training, the best-performing control field $\Omega_p(t)$ and $\Omega_s(t)$, and the corresponding state vector $\mathbf{x}(t)$, are stored when the total loss reaches a new minimum. The control area is calculated using numerical integration. The final results demonstrate that the control pulses generated by the PINNs framework enable efficient nuclear state population transfer, and the learned control fields are further verified through numerical computation.
\section{HIGHLY EFFICIENT NUCLEAR COHERENT POPULATION TRANSFER} \label{section3}
\begin{figure}[t]
\centering
\includegraphics[width=\columnwidth]{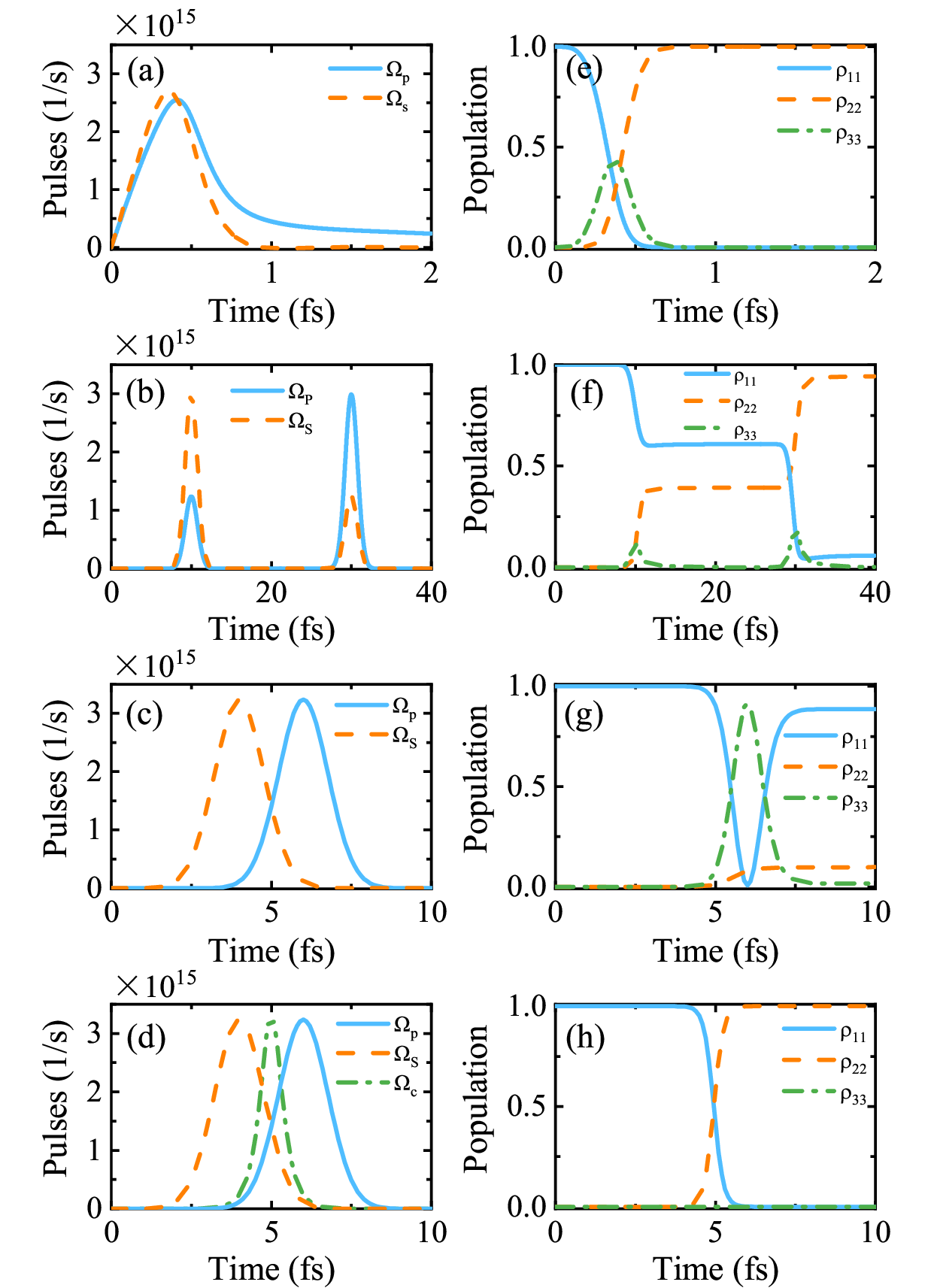}
\caption{Dynamical comparison of PINNs and traditional methods in the nuclear three-level system of $^{172}$Yb with short excited state lifetime. (a)-(d) Pulse sequences $\Omega_{p,s,c}(t)$ for each control method: PINNs, coincident pulses, STIRAP and MIE. (e)-(h) Corresponding population evolutions $\rho_{ii}(t)$ during the state transfer.}
\label{fig.3}
\end{figure}

To systematically analyze the performance of different quantum control methods in nuclear systems with significantly different lifetimes, we consider two nuclear systems: $^{172}$Yb with a lifetime of $11$ fs and $^{229}$Th with a lifetime of $0.172$ ns. Their additional characteristic parameters are given in Appendix \ref{appendix2}. For comparison, three representative traditional quantum control approaches are selected: coincident pulse technique, STIRAP and MIE. Our study focuses on three key metrics: (i) population transfer efficiency $P_2=\rho_{22}$, (ii) the shortest control time $t_f$ required to achieve the target transfer efficiency, (iii) the pulse area $A = \int_0^{t_f} \Omega(t) \, dt$. All control methods drive the $\vert 1\rangle \leftrightarrow \vert 3\rangle$ and $\vert 2\rangle \leftrightarrow \vert 3\rangle$ transitions via Rabi frequency, aiming to realize a high efficiency $\vert 1\rangle \rightarrow \vert 2\rangle$ population transfer while suppressing the intermediate-state occupation. To ensure a fair comparison, all Rabi frequency amplitudes are constrained by a maximum Rabi frequency $\Omega_{s,p} \leq 5$.
\begin{figure}[t]
\centering
\includegraphics[width=\columnwidth]{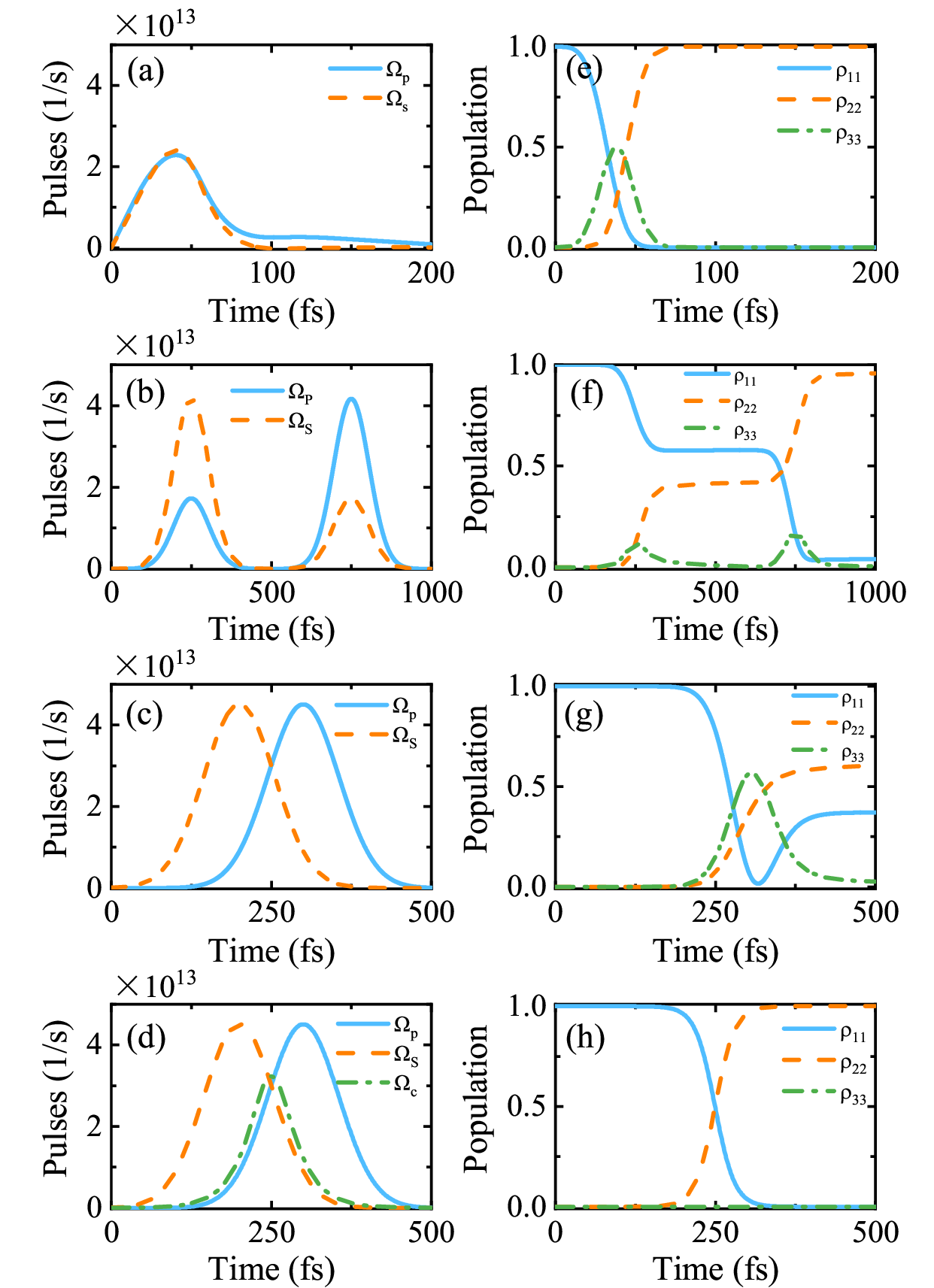}
\caption{Dynamical comparison of PINNs and traditional methods in the nuclear three-level system of $^{229}$Th with long excited state lifetime. (a)-(d) Pulse sequences $\Omega_{p,s,c}(t)$ for each control method: PINNs, coincident pulses, STIRAP and MIE. (e)-(h) Corresponding population evolutions $\rho_{ii}(t)$ during the state transfer.}
\label{fig.4}
\end{figure}

We calculate the time evolution of the target state for different nuclei using four different methods.
Figure~\ref{fig.3} presents the pulses generated by PINNs and the three traditional methods, along with the corresponding population dynamics in the short-lifetime $^{172}$Yb system.
The results for transfer efficiency, control time, and pulse area across different schemes are reported in Table~\ref{Table1}.
Due to the femtosecond-scale excited-state lifetime, the population transfer must be completed within an extremely short duration while minimizing the population in the intermediate state to suppress non-radiative losses and decoherence.
Figure~\ref{fig.3} (a) and (e) illustrate the PINNs approach employs a neural network embedded with physical constraints to autonomously generate optimized pulse sequences. It achieves nearly perfect population transfer of $0.9999$ within only $2$~fs, using a small pulse area of $2.18$.
\begin{table}[t]
\centering
\caption{Comparison of four control methods applied to the $^{172}$Yb and $^{229}$Th nuclear systems. The evaluation focuses on three aspects: the final population in state $|2\rangle$ denoted by $P_2$, the pulse area, defined as $A = \int_0^{t_f} dt\, \sqrt{|\Omega_p(t)|^2 + |\Omega_s(t)|^2}$, and the total control time $t_f$ (in fs).}
\small
\setlength{\tabcolsep}{2.5pt} 
\renewcommand{\arraystretch}{1.3} 
\begin{tabular}{lccccc}
\hline\hline
Nucleus & Performance & PINNs & Coin. Pulses & STIRAP & MIE \\
\hline
\multirow{3}{*}{$^{172}$Yb}
    & $P_2$   & 0.9999  & 0.9422   & 0.0979  & 0.9999 \\
    & $A$     & 2.11  & 12.56  & 11.57 & 12.74 \\
    & $t_f$   & 2     & 40     & 10    & 10 \\
\hline
\multirow{3}{*}{$^{229}$Th}
    & $P_2$   & 0.9999  & 0.9569  & 0.6044   & 0.9999 \\
    & $A$     & 1.79  & 12.56 & 10.75  & 11.43 \\
    & $t_f$   & 200   & 1000  & 500    & 500 \\
\hline\hline
\end{tabular}
\label{Table1}
\end{table}
In comparison, the coincident pulses scheme, depicted in Fig.~\ref{fig.3} (b) and (f), achieves population transfer through a train of coincident pulses. To suppress the population in the excited state, a mixing angle \(\phi_k = (2k - 1)\pi/4N\) \((k = 1, 2, \dots, N)\)
is introduced as the control parameter for each pair of pulses, where each pulse has a pulse area of $2\pi$~\cite{PhysRevC.94.054601}. Here we take $N=2$. Based on these parameters, the required laser intensities for each step are calculated in the $^{172}\mathrm{Yb}$ nuclear system. Although increasing the number of pulse pairs can suppress laser intensity fluctuation, the population transfer process requires up to $40$ fs, and the relatively large pulse area entails higher energy consumption.
The STIRAP protocol, shown in Fig.~\ref{fig.3} (c) and (g), employs counterintuitive pulse sequences to achieve population transfer and offers robustness to parameter variations~\cite{PhysRevA.102.023715}. Moreover, this method relies on slow adiabatic evolution, which is fundamentally incompatible with the short excited-state lifetime of the $^{172}$Yb nucleus. As a result, adiabatic pathways cannot be established, thereby limiting the transfer efficiency.
The MIE approach, illustrated in Fig.~\ref{fig.3} (d) and (h), achieves high population transfer efficiency within $10$~fs by introducing an additional control field $\Omega_c$ that directly couples states $\vert1\rangle$ and $\vert2\rangle$, effectively suppressing intermediate-state occupation and reducing decoherence. This method requires precise femtosecond-scale timing and narrow bandwidth, posing significant experimental challenges. In addition, the total pulse area is approximately six times large as that in the PINNs approach, leading to significantly higher energy consumption.
In contrast to the limitations of traditional methods, the PINNs approach demonstrates a remarkable advantage in the ultrafast regime. It autonomously generates optimized pulse sequences that compress both control time and pulse area, which is essential for achieving high-fidelity control in short-lifetime systems.

The corresponding results for the long-lifetime nucleus $^{229}$Th are presented in Fig.~\ref{fig.4}. It shows that the PINNs method remains efficient, generating optimized pulses similar to those in the $^{172}\mathrm{Yb}$ system and achieving complete transfer within $200$~fs. While coincident pulses in Fig.~\ref{fig.4} (b) and STIRAP in Fig.~\ref{fig.4} (c) achieve population transfer due to relaxed time constraints, they still suffer from intermediate-state population and require larger pulse areas as well as hundreds of femtoseconds in control time. MIE in Fig.~\ref{fig.4} (d) achieves fast and efficient population transfer with minimal intermediate-state occupation, but it requires a relatively high pulse area. Compared with the nucleus $^{172}$Yb with a short excited-state lifetime, the nucleus $^{229}$Th with a long excited-state lifetime requires a longer time to achieve efficient nuclear state transition.
Table \ref{Table1} also summarizes the performance metrics obtained for population transfer in the $^{229}$Th nuclear system, including the transfer efficiency, control time and pulse area.
Notably, for PINNs, the long lifetime of $^{229}$Th relaxes the timing constraint, enabling the neural network to generate optimized pulses with peak Rabi frequencies lower than those employed in traditional schemes, which decreases the required laser intensity. Furthermore, for both short and long-lifetime nuclei, PINNs accomplish efficient population transfer before the total control time, within $1~\mathrm{fs}$ for the short-lifetime nucleus and within $100~\mathrm{fs}$ for the long-lifetime nucleus. These results indicate that PINNs achieve efficient population transfer and demonstrate clear advantages over conventional methods.

\section{CONCLUSION} \label{section4}
In summary, we have investigated population transfer in a three-level nuclear system driven by laser pulses optimized via PINNs. By systematically comparing the PINNs approach with three representative conventional control strategies, namely STIRAP, coincident pulses and the MIE method, we have demonstrated its superior performance in both short and long excited-state lifetime nuclear systems. Taking $^{172}$Yb and $^{229}$Th as representative nuclei, the PINNs scheme achieves nearly $100\%$ population transfer, with higher efficiency, shorter control time, smaller pulse area, and the infidelity reaches the level of $10^{-4}$.
These results validate the effectiveness of the PINNs approach in nuclear state control and provide an efficient control scheme for laser-driven nuclear systems with diverse lifetimes.
Such efficient population transfer plays a pivotal role in enabling the precise control of nuclei, which is essential for ensuring the stability and accuracy of next-generation nuclear clocks based on the long-lifetime isomeric state of $^{229}$Th~\cite{Seiferle2019,kraemer2023observation,zhang2024frequency,PhysRevLett.134.113801}.
It is also crucial for the performance of nuclear batteries, where efficient and clean energy conversion relies on controlled excitation and triggered decay of isomeric nuclear states~\cite{PRELAS2014117,li2024micronuclear}. With the advent of XFEL, coherent nuclear population transfer becomes feasible, opening new directions in nuclear photonics, precision nuclear spectroscopy, and hybrid quantum systems bridging atomic and nuclear physics~\cite{RevModPhys.84.1177,Liao2014}.
The integration of machine learning-based optimization with quantum dynamics holds great potential for future applications such as ultra-precise timekeeping~\cite{von2016direct}, controlled nuclear energy release~\cite{PhysRevLett.99.172502}, and high-performance quantum batteries~\cite{dou2022highly,PhysRevB.105.115405,PhysRevA.109.032201,PhysRevB.109.235432,RevModPhys.96.031001}.
\section*{Acknowledgments}
The work is supported by the National Natural Science Foundation of China (Grant No. 12475026) and the Natural Science Foundation of Gansu Province (No. 25JRRA799).
\appendix
\section{THE DETAILS OF PINNs}\label{appendix1}
In this appendix, we provide a detailed derivation of the real-valued equations and the system matrix $A(\lambda, u(t))$, reformulating the master equation for direct use in the PINNs framework introduced in the main text.

The density matrix of the three-level nuclear system can be explicitly written as
\begin{eqnarray}\label{rho explicit form}
\rho(t)=\begin{pmatrix}
\rho_{11} & \rho_{12} & \rho_{13} \\
\rho_{21} & \rho_{22} & \rho_{23} \\
\rho_{31} & \rho_{32} & \rho_{33}
\end{pmatrix}.
\end{eqnarray}
Substituting the Hamiltonian Eqs.~(\ref{Hamiltonian}) and (\ref{decoherence matrix}) into the nuclear master equation Eq.~(\ref{nuclear master equation}), we obtain the following set of differential equations governing the evolution of the density matrix elements:
\begin{eqnarray}\label{explicit form}
\begin{aligned}
\partial_{t} \rho_{11} & =\frac{i}{2}\Omega_{p}\rho_{13}-\frac{i}{2}\Omega_{p}\rho_{31}+\Gamma B_{31}\rho_{33},\\
\partial_{t} \rho_{22} & =\frac{i}{2}\Omega_{s}\rho_{23}-\frac{i}{2}\Omega_{s}\rho_{32}+\Gamma B_{32}\rho_{33},\\
\partial_{t} \rho_{33} & =-\frac{i}{2}\Omega_{p}\rho_{13}+\frac{i}{2}\Omega_{p}\rho_{31}-\frac{i}{2}\Omega_{s}\rho_{23}+\frac{i}{2}\Omega_{s}\rho_{32}-\Gamma \rho_{33},\\
\partial_{t} \rho_{12} & =\frac{i}{2}\Omega_{s}\rho_{13}-\frac{i}{2}\Omega_{p}\rho_{32},\\
\partial_{t} \rho_{21} & =-\frac{i}{2}\Omega_{s}\rho_{31}+\frac{i}{2}\Omega_{p}\rho_{23},\\
\partial_{t} \rho_{13} & =\frac{i}{2}\Omega_{p}\rho_{11}+\frac{i}{2}\Omega_{s}\rho_{12}-\frac{i}{2}\Omega_{p}\rho_{33}-\frac{1}{2}\Gamma \rho_{13},\\
\partial_{t} \rho_{31} & = -\frac{i}{2}\Omega_{p}\rho_{11}-\frac{i}{2}\Omega_{s}\rho_{21}+\frac{i}{2}\Omega_{p}\rho_{33}-\frac{1}{2}\Gamma \rho_{31},\\
\partial_{t} \rho_{23} & =\frac{i}{2}\Omega_{p}\rho_{21}+\frac{i}{2}\Omega_{s}\rho_{22}-\frac{i}{2}\Omega_{s}\rho_{33}-\frac{1}{2}\Gamma \rho_{23},\\
\partial_{t} \rho_{32} & =-\frac{i}{2}\Omega_{p}\rho_{12}-\frac{i}{2}\Omega_{s}\rho_{22}+\frac{i}{2}\Omega_{s}\rho_{33}-\frac{1}{2}\Gamma \rho_{32}.\\
\end{aligned}
\end{eqnarray}
Here, the density matrix elements are decomposed into their real and imaginary parts. Since \(\rho_{ij}\) and \(\rho_{ji}\) are complex conjugates, we define:
\begin{equation} \label{Density Matrix Decomposition}
\begin{array}{ll}
\text{Re}[\rho_{12}] = \dfrac{\rho_{12} + \rho_{21}}{2}, & \text{Im}[\rho_{12}] = \dfrac{\rho_{12} - \rho_{21}}{2i}, \\
\text{Re}[\rho_{13}] = \dfrac{\rho_{13} + \rho_{31}}{2}, & \text{Im}[\rho_{13}] = \dfrac{\rho_{13} - \rho_{31}}{2i}, \\
\text{Re}[\rho_{23}] = \dfrac{\rho_{23} + \rho_{32}}{2}, & \text{Im}[\rho_{23}] = \dfrac{\rho_{23} - \rho_{32}}{2i}.
\end{array}
\end{equation}
We now introduce a new set of real-valued variables $x_i$ to represent the populations and the real and imaginary parts of coherences. By applying the Eq.~(\ref{Density Matrix Decomposition}) to the master equation in Eq.~(\ref{explicit form}), we obtain the following nine real-valued differential equations:
\begin{equation}\label{eq:Zeqs}
\begin{aligned}
0 &= \dot{x}_1 - \Gamma B_{31} {x}_3 + \Omega_p x_7, \\
0 &= \dot{x}_2 - \Gamma B_{32} {x}_3 + \Omega_s x_9, \\
0 &= \dot{x}_3 + \Gamma {x}_3 - \Omega_p x_7 - \Omega_s x_9, \\
0 &= \dot{x}_4 + \frac{\Omega_s}{2} x_7 + \frac{\Omega_p}{2} x_9, \\
0 &= \dot{x}_5 - \frac{\Omega_s}{2} x_6 + \frac{\Omega_p}{2} x_8, \\
0 &= \dot{x}_6 + \frac{\Omega_s}{2} x_5 + \frac{\Gamma}{2} x_6, \\
0 &= \dot{x}_7 - \frac{\Omega_p}{2} x_1 + \frac{\Omega_p}{2} x_3 - \frac{\Omega_s}{2} x_4 + \frac{\Gamma}{2} x_7,\\
0 &= \dot{x}_8 - \frac{\Omega_p}{2} x_5 + \frac{\Gamma}{2} x_8, \\
0 &= \dot{x}_9 - \frac{\Omega_s}{2} x_2 + \frac{\Omega_s}{2} x_3 - \frac{\Omega_p}{2} x_4 + \frac{\Gamma}{2} x_9.  \\
\end{aligned}
\end{equation}
During numerical computation, this system can be compactly expressed in matrix form as the following \(9 \times 9\) matrix \(A(\lambda, u(t))\):
\begin{equation}\label{eq:matrix}
\renewcommand{\arraystretch}{1.6}
\left(
\begin{array}{ccccccccc}
0 & 0 & \Gamma B_{31} & 0 & 0 & 0 & -\Omega_p & 0 & 0 \\
0 & 0 & \Gamma B_{32} & 0 & 0 & 0 & 0 & 0 & -\Omega_s \\
0 & 0 & -\Gamma & 0 & 0 & 0 & \Omega_p & 0 & \Omega_s \\
0 & 0 & 0 & 0 & 0 & 0 & -\frac{\Omega_s}{2} & 0 & -\frac{\Omega_p}{2} \\
0 & 0 & 0 & 0 & 0 & \frac{\Omega_s}{2} & 0 & -\frac{\Omega_p}{2} & 0 \\
0 & 0 & 0 & 0 & -\frac{\Omega_s}{2} & -\frac{\Gamma}{2} & 0 & 0 & 0 \\
\frac{\Omega_p}{2} & 0 & -\frac{\Omega_p}{2} & \frac{\Omega_s}{2} & 0 & 0 & -\frac{\Gamma}{2} & 0 & 0 \\
0 & 0 & 0 & 0 & \frac{\Omega_p}{2} & 0 & 0 & -\frac{\Gamma}{2} & 0 \\
0 & \frac{\Omega_s}{2} & -\frac{\Omega_s}{2} & \frac{\Omega_p}{2} & 0 & 0 & 0 & 0 & -\frac{\Gamma}{2}
\end{array}
\right)
\end{equation}
This compact matrix representation is well-suited for PINNs-based optimization, in which the control fields $\Omega_p(t)$ and $\Omega_s(t)$ are learned to satisfy the dynamical constraints while steering the nuclear system toward the desired target state.
\begin{widetext}
\section{THE CHARACTERISTIC PARAMETERS FOR NUCLEI}\label{appendix2}
This appendix provides the characteristic physical parameters of the $^{172}$Yb and $^{229}$Th nuclear systems used in our study.
\renewcommand{\tabcolsep}{0.09cm}
\renewcommand{\arraystretch}{1.8}
\begin{table*}[htbp]
\renewcommand{\arraystretch}{1.5}
  \centering
  \caption{The characteristic parameters for nuclei. $E_{i}$ is the energy of state $|i\rangle$ $(i=1,2,3)$ (in keV). $\gamma$ denotes relativistic factor. $\Gamma$ is the linewidth of state $|3\rangle$ (in meV). $B_{31}$ and $B_{32}$ are branching ratios of $|3\rangle \to |i\rangle (i=1,2)$, respectively. $\mu L_{ij}$, $\mu\in\{E,M\}$ are the multipolarities and $\mathbb{B}_{ij}(\mu L_{ij})$ (in Weisskopf units, wsu) denote the reduced matrix elements for the transitions $|i\rangle \to |3\rangle (i=1,2)$. The peak intensities of pump and Stokes laser pulse (in W/cm$^{2}$) have been given in Ref. \cite{PhysRevC.87.054609,PhysRevC.94.054601,PhysRevC.110.034606,NNDC2011}.}
  \scalebox{0.88}{
      \begin{tabular}{cccccccccccccccc}
      \hline\hline
       Nucleus &$E_{3}$ &$E_{2}$ &$E_{1}$ &$\gamma$ &$\Gamma$  &\multicolumn{2}{c}{Branching ratio} &\multicolumn{2}{c}{Multipolarity} &$\mathbb{B}_{13}$ &$\mathbb{B}_{23}$ &$\Omega_{p0}$ &$\Omega_{s0}$ &$I_{p}$ $\times$10$^{21}$ &$I_{s}$ $\times$10$^{21}$ \\
       \textit{} &(keV) &(keV) &(keV) &\textit{} &\textit{} (meV) &$B_{31}$ &$B_{32}$ &$L_{13}$ &$L_{23}$ &(wsu) &(wsu) &(1/s) &(1/s) &(W/cm$^{2}$) &(W/cm$^{2}$)\\
       \hline
      $^{229}$Th &29.19 &0.008355 &0.00 &1.18 &5.45 &0.0936 &0.9250 &$M1$ &$E2$ &0.003 &44.9 &$2.7174$ $\times$10$^{3}$ &$8.5621$ $\times$10$^{3}$  &$0.2750$ &$0.0277$ \\
      \hline
      $^{172}$Yb &1599.87 &78.74 &0.00 &64.5 &42.50 &0.3911 &0.6017 &$E1$ &$E1$  &0.0018 &12.3 &$7.6374$ $\times$10$^{4}$ &$1.4114$ $\times$10$^{5}$ &$1.7923$ &$0.5248$ \\
      \hline\hline
      \end{tabular}
      }
  \label{TAB.2}
\end{table*}
\end{widetext}

\bibliography{refercence}

\end{document}